\documentclass[twocolumn,showpacs,preprintnumbers,amsmath,amssymb,superscriptaddress]{revtex4}

\usepackage{graphicx}
\usepackage{epstopdf}
\usepackage{dcolumn}
\usepackage{bm}

\begin{document}

\title{Ultraslow fluctuations in the pseudogap states of HgBa$_{2}$CaCu$_{2}$O$_{6+\delta}$} 

\author{Yutaka Itoh\thanks{E-mail:yitoh@cc.kyoto-su.ac.jp}}
 \affiliation{Department of Physics, Graduate School of Science, Kyoto Sangyo University, Kamigamo-Motoyama, Kika-ku, Kyoto 603-8555, Japan}
\author{Takato Machi }
 \affiliation{AIST Tsukuba East, Research Institute for Energy Conservation, 1-2-1 Namiki, Tsukuba, Ibaraki 305-8564, Japan}
\author{Ayako Yamamoto}
 \affiliation{Graduate School of Engineering and Science, Shibaura Institute of Technology, 3-7-5 Toyosu, Koto-ku, Tokyo 135-8548, Japan}

\date{\today}

\begin{abstract}
We report the transverse relaxation rates 1/$T_2$'s of the $^{63}$Cu nuclear spin-echo envelope for double-layer high-$T_\mathrm{c}$ cuprate superconductors HgBa$_{2}$CaCu$_{2}$O$_{6+\delta}$ from underdoped to overdoped.  
The relaxation rate 1/$T_\mathrm{2L}$ of the exponential function (Lorentzian component) shows a peak  at 220$-$240 K in the underdoped ($T_\mathrm{c}$ = 103 K) and the optimally doped ($T_\mathrm{c}$ = 127 K) samples but no peak in the overdoped ($T_\mathrm{c}$ = 93 K) sample. 
The enhancement in 1/$T_\mathrm{2L}$ suggests development of the zero frequency components of local field fluctuations. 
Ultraslow fluctuations are hidden in the pseudogap states. 
\end{abstract}



\maketitle

\section{Introduction}
Normal state precursory diamagnetism~\cite{Iguchi,Ong1,Ong2}, a magnetic field induced charge-stripe order~\cite{Julien} and a short-range charge density wave order~\cite{CDW,RIX} have renewed our interests in the pseudogap states of high-$T_\mathrm{c}$ cuprate superconductors. 
The nature of the pseudogap state has been one of the central issues in studying strong correlation effects.  

We focus on the transverse relaxation of nuclear spins in the system. 
For the typical high-$T_\mathrm{c}$ cuprate superconductor YBa$_2$Cu$_3$O$_7$ in a static magnetic field along $c$ axis, the plane-site Cu nuclear spin-echo transverse relaxation 
is caused by the indirect nuclear spin-spin coupling via the electron spin fluctuations and the nuclear spin-lattice relaxation process due to the electron spin fluctuations~\cite{P,PC}. 
The static nuclear spin-spin coupling leads to a Gaussian relaxation function with a time constant $T_\mathrm{2G}$, that is a spin-echo amplitude $E(t)$ at time $t$ as $E(t)$ $\propto$ ${\rm exp}[-{\frac{1}{2}}(t/T_\mathrm{2G})^2]$~\cite{P,PC}.
The electron spin fluctuations lead to a single exponential relaxation function with a time constant $T_\mathrm{2}$ ($T_1$ process) associated with a nuclear spin-lattice relaxation time $T_1$ using the Redfield theory, that is $E(t)$ $\propto$ ${\rm exp}(-t/T_2)$~\cite{P,Slichter}. 
The Cu nuclear spin-echo transverse relaxation curve is expressed by the product of the exponential and Gaussian functions in the static limit of $T_\mathrm{1}{\gg}T_\mathrm{2G}$, that is $E(t)$ = $E(0){\rm exp}[-t/T_2-{\frac{1}{2}}(t/T_\mathrm{2G})^2]$.

The nuclear spin-lattice relaxation rate divided by temperature 1/$T_1T$ is associated with the low frequency part of the dynamical spin susceptibility $\chi^{\prime\prime}$({\mbox{\boldmath $q$}}, $\omega$) ({\mbox{\boldmath $q$}} is a wave vector, and $\omega$ is a frequency) of unpaired electrons~\cite{Moriya}, while the Gaussian relaxation rate 1/$T_\mathrm{2G}$ is associated with the static staggered spin susceptibility $\chi^{\prime}$({\mbox{\boldmath $Q$}}) ({\mbox{\boldmath $Q$}} is the antiferromagnetic wave vector)~\cite{Moriya2,PC}. 
$\chi^{\prime}$({\mbox{\boldmath $Q$}}) is related to $\chi^{\prime\prime}$({\mbox{\boldmath $Q$}}, $\omega$) through the Kramers-Kronig relation. 
The temperature dependence of $\chi^{\prime}$({\mbox{\boldmath $Q$}}) through 1/$T_\mathrm{2G}$ for the high-$T_\mathrm{c}$ cuprate superconductors tells us the $d_{x^2-y^2}$ wave pairing symmetry in the superconducting state~\cite{Itoh}, the self-consistently renormalized spin fluctuation effects on the Curie-Weiss type antiferromagnetic correlation length~\cite{ItohSCR}, and the pseudo spin-gap spectrum~\cite{Itoh,ItohHg1212}. 
Dynamic scaling laws on $\chi^{\prime\prime}$({\mbox{\boldmath $q$}}, $\omega$) for low dimensional antiferromagnetic Heisenberg systems have been tested by the measurements of 1/$T_\mathrm{2G}$~\cite{Imai,Takigawa}. 
Thus, the Gaussian decay rate 1/$T_\mathrm{2G}$ has provided us with rich information on the microscopic properties of the correlated materials. 
In this paper, however, our main concern is a transverse relaxation time $T_2$ due to $T_1$ process (Lorentzian component) but not the Gaussian decay time $T_\mathrm{2G}$. 

To be exact, a longitudinal nuclear (spin-lattice) relaxation time $T_1$ and a transverse relaxation time $T_2$ ($T_1$ process) are caused by different frequency parts of the local field fluctuations ~\cite{Moriya}. 
For convenience, let us assume a nuclear spin $I$ = 1/2. 
$T_1$ probes the transverse fluctuation $J_{\perp}(\nu_n)$ at a Larmor frequency $\nu_n$ as 1/$T_1 = 2J_{\perp}(\nu_n)$, while $T_2$ probes the additional longitudinal fluctuation at the zero frequency as 1/$T_2$ = 1/2$T_1$ + $J_{\parallel}(0)$~\cite{Moriya,Jac,Slichter}.
Zero frequency fluctuations represent ultraslow dynamics of the system, e.g., glassy nature and frustration effects. 
Glassy charge-spin stripe orderings with ultraslow fluctuations and wipeout effects have been observed by nuclear quadrupole resonance (NQR) measurements for the La-based 214 family~\cite{LSCO1,LSCO2,LSCO3}.

If the longitudinal fluctuation at the zero frequency is equal to that at the Larmor frequency as $J_{\parallel}(0) = J_{\parallel}(\nu)$, $T_2$ can be estimated from $T_1$ using the Redfield theory~\cite{Slichter}.
The $T_2$ estimated from the experimental $T_1$ is denoted by $T_\mathrm{2R}$ after Ref.~\cite{P}.
The experimental $T_2$ in the exponential time development of the transverse relaxation function is denoted by $T_\mathrm{2L}$. 

HgBa$_{2}$CaCu$_{2}$O$_{6+\delta}$ (Hg1212) is a duble-CuO$_2$-layer system of the high-$T_\mathrm{c}$ cuprate superconductors~\cite{Antipov}.
The optimized $T_\mathrm{c}$ of about 127 K is the highest among the ever reported double-layer cuprate superconductors,
which is associated with the flatness of the CuO$_2$ planes in a unit cell~\cite{Antipov,Fukuoka}. 
Not only high $T_\mathrm{c}$ but also large scale pseudo spin-gap turned out to characterize Hg1212 even at the optimally doping level through Cu NMR studies~\cite{ItohHg1212}.

In this paper, we report the $^{63}$Cu nuclear spin-echo transverse relaxation rates for Hg1212 with $T_\mathrm{c}$ = 103 K at the underdoping level, 127 K at the optimally doping level, and 93 K at the overdoping level.
We found the strong enhancement in the transverse relaxation rate 1/$T_\mathrm{2L}$ of the exponential function at 220$-$240 K for the underdoped and optimally doped samples, which could not be explained by anisotropic $^{63}$Cu nuclear spin-lattice relaxation times.
The peak in 1/$T_\mathrm{2L}$ at 220$-$240 K suggests the emergence of novel zero frequency fluctuations in the pseudogap states.  
 
\section{Experiments}
NMR experiments were performed for magnetically $c$ axis aligned powder samples of Hg1212.
The samples were prepared and characterized in Ref.~\cite{Fukuoka}. 
A phase-coherent-type pulsed spectrometer was utilized to perform the $^{63}$Cu NMR (nuclear spin $I$ = 3/2) experiments at 85.2 MHz ($\sim$7.45 T).
Most of the $^{63}$Cu NMR results have already been published in Ref.~\cite{ItohHg1212} except the present $T_2$ data.      
The time development of the $^{63}$Cu nuclear spin-echo envelope was measured by recording the spin-echo amplitude $E$(2$\tau$) following a sequence of $\pi$/2$-$$\tau$$-$$\pi$ pulses.

\begin{figure}[b]
 \begin{center} 
 \includegraphics[width=1.03\linewidth]{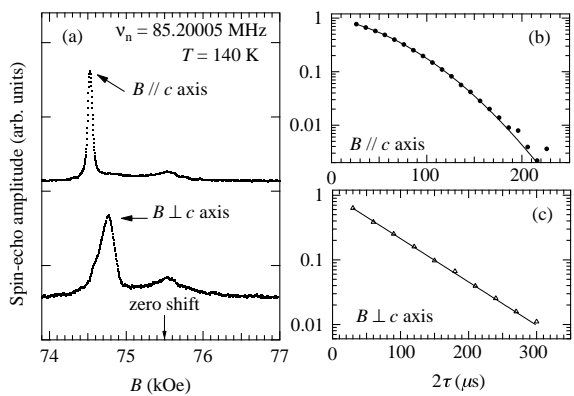}
 \end{center}
 \caption{\label{f1}
(a) Central transition lines  ($I_z$ = 1/2 $\leftrightarrow$ -1/2) of field-swept quadrupole-split $^{63}$Cu NMR spectra ($B$$\parallel$$c$ axis and $B$$\perp$$c$ axis) for the optimally doped Hg1212 ($T_\mathrm{c}$ = 127 K) at 140 K and 85.2 MHz.
Transverse relaxation curves  $E$(2$\tau$) of the $^{63}$Cu nuclear spin-echoes for $B$$\parallel$$c$ axis (b) and $B$$\perp$$c$ axis (c). 
 }
 \end{figure}
 
\begin{figure}[h]
 \begin{center} 
 \includegraphics[width=0.76\linewidth]{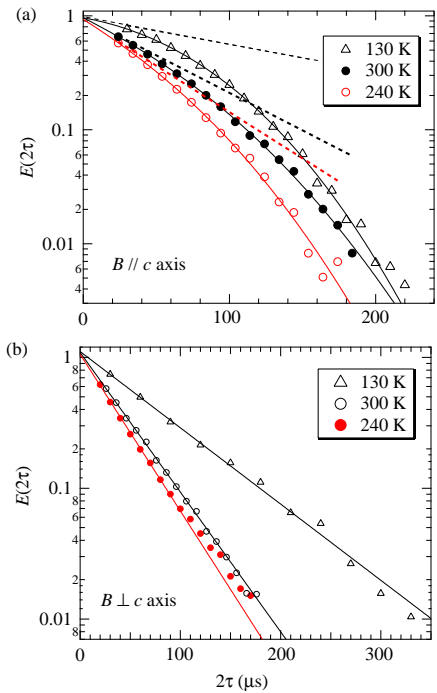}
 \end{center}
 \caption{\label{f2}(Color online)
(a) Transverse relaxation curves $E$(2$\tau$) of the $^{63}$Cu nuclear spin-echo for the optimally doped Hg1212 ($T_\mathrm{c}$ = 127 K) at 130, 240 and 300 K ($B$$\parallel$$c$ axis). 
Solid curves are the least-squares fitting results using Eq.~(\ref{eqSEO}).  
For visual guides, dashed lines indicate single exponential functions with $T_\mathrm{2L}$ isolated from Eq.~(\ref{eqSEO}). 
(b) Transverse relaxation curves $E$(2$\tau$) of the $^{63}$Cu nuclear spin-echo at 130, 240 and 300 K ($B$$\perp$$c$ axis). 
Solid lines are the least-squares fitting results using a single exponential function.  
 }
 \end{figure}
  
The $^{63}$Cu nuclear spin-echo transverse relaxation curves $E$(2$\tau$) were analyzed by  
\begin{equation}
E(2\tau) = E(0){\rm exp}\left[-\frac{2\tau}{T_\mathrm{2L}} - \frac{1}{2}\left(\frac{2\tau}{T_\mathrm{2G}}\right)^2\right]f(2\tau),
\label{eqSEO}
\end{equation}  
where $E$(0), $T_\mathrm{2L}$, and $T_\mathrm{2G}$ are the fitting parameters~\cite{P}. 
$f(2\tau)$ is the correction function of the $I_z$ fluctuation effects and ln$f(2\tau)$ consists of the higher order terms of $\tau^3$ and $\tau^4$ given in Refs.~\cite{Curro,CS}. 
The Gaussian decay rate 1/$T_\mathrm{2G}$ is a measure of the indirect nuclear spin-spin coupling constant~\cite{Moriya2,P,PC}. 
The present Gaussian decay rates 1/$T_\mathrm{2G}$'s were estimated by including the correction of the $I_z$ fluctuation effect.
The uncorrected 1/$T_\mathrm{2G}$'s, being overestimated, have been reported in Ref.~\cite{ItohHg1212}.
It is the first time to report 1/$T_\mathrm{2L}$.  
 
Figure~\ref{f1} (a) shows the central transition lines  ($I_z$ = 1/2 $\leftrightarrow$ -1/2) of the field-swept quadrupole-split $^{63}$Cu NMR spectra in an external magnetic field $B$ along the oriented $c$ axis ($B$$\parallel$$c$) and perpendicular to the $c$ axis ($B$$\perp$$c$) for the optimally doped Hg1212 ($T_\mathrm{c}$ = 127 K) at 140 K and 85.2 MHz.
The full width of the half maximum of the central transition line is about 76 G for $B$$\parallel$$c$ axis and about 230 G for $B$$\perp$$c$ axis.
The $\pi$/2 pulse of $H_1$ $\sim$ 100 G can excite all the relevant nuclear spins for $B$$\parallel$$c$ axis.
No $H_1$ dependence of the spin-echo decay was also observed for $B$$\perp$$c$ axis.  

Figures~\ref{f1} (b) and (c) show the $^{63}$Cu nuclear spin-echo transverse relaxation curves $E$(2$\tau$) of the central transition lines for $B$$\parallel$$c$ axis (b) and $B$$\perp$$c$ axis (c). 
Solid curves are the least-squares fitting results using Eq.~(\ref{eqSEO}).  
The Gaussian component for $B$$\perp$$c$ axis was negligible, which indicates a weak coupling constant of $I_{\pm}I_{\mp}$ (the in-plane components of the nuclear spin $I$) and no contribution from the nonsecular mutual spin flip terms~\cite{ImaiT2,Itoh2}. 

Figures~\ref{f2} shows the temperature dependence of the $^{63}$Cu nuclear spin-echo transverse relaxation curve $E$(2$\tau$) for $B$$\parallel$$c$ axis (a) and for $B$$\perp$$c$ axis (b). 
In Fig.~\ref{f2} (a), solid curves are the least-squares fitting results using Eq.~(\ref{eqSEO}). 
For $B$$\parallel$$c$ axis, the initial decay of $E$(2$\tau$) at 240 K is faster than those at 130 K and 300 K, which is shown by the dashed lines. 
In Fig.~\ref{f2} (b), solid lines are the least-squares fitting results using a single exponential function. 
For $B$$\perp$$c$ axis, the exponential decay of $E$(2$\tau$) at 240 K is faster than those at 130 K and 300 K. 
 
\section{Transverse relaxation rates}
\begin{figure}[b]
 \begin{center}
 \includegraphics[width=0.80\linewidth]{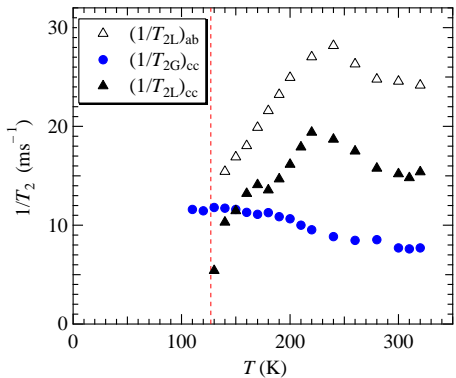}
 \end{center}
 \caption{\label{f3}(Color online)
Transverse relaxation rates (1/$T_\mathrm{2L}$)$_{cc}$, (1/$T_\mathrm{2G}$)$_{cc}$, and (1/$T_\mathrm{2L}$)$_{ab}$ of the $^{63}$Cu nuclear spin-echo for the optimally doped Hg1212 ($T_\mathrm{c}$ = 127 K). 
The dashed line indicates $T_\mathrm{c}$ = 127 K.
 }
 \end{figure} 
 
 \begin{figure}[h]
 \begin{center}
 \includegraphics[width=0.80\linewidth]{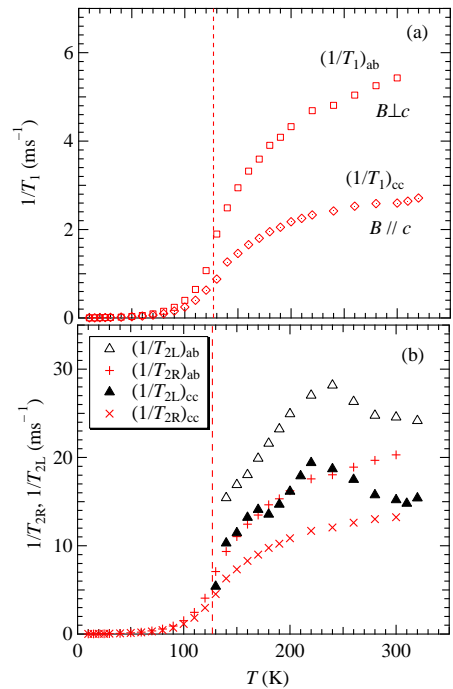}
 \end{center}
 \caption{\label{f4}(Color online)
(a) $^{63}$Cu nuclear spin-lattice relaxation rates $(1/T_{1})_{cc}$ and $(1/T_{1})_{ab}$ for the optimally doped Hg1212 ($T_\mathrm{c}$ = 127 K).
(b) $^{63}$Cu nuclear spin-echo transverse relaxation rates $(1/T_\mathrm{2L})_{cc}$ and $(1/T_\mathrm{2L})_{ab}$. 
$(1/T_\mathrm{2R})_{cc}$ and $(1/T_\mathrm{2R})_{cc}$ were estimated using the Redfield theory~\cite{Slichter} from the anisotropic $T_1$'s in (a). 
The dashed line indicates $T_\mathrm{c}$ = 127 K.
 }
 \end{figure}
 
\begin{figure}[h]
 \begin{center}
 \includegraphics[width=0.80\linewidth]{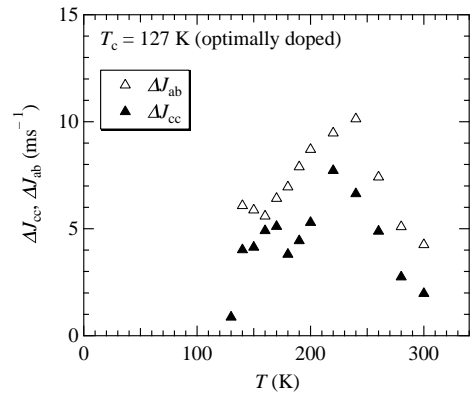}
 \end{center}
 \caption{\label{f5}
$\Delta J_{cc}$ and $\Delta J_{ab}$ against temperature for the optimally doped Hg1212. 
}
 \end{figure}
 
Figure~\ref{f3} shows $^{63}$Cu nuclear spin-echo transverse relaxation rates (1/$T_\mathrm{2L}$)$_{cc}$, (1/$T_\mathrm{2G}$)$_{cc}$, and (1/$T_\mathrm{2L}$)$_{ab}$ for the optimally doped Hg1212 ($T_\mathrm{c}$ = 127 K). 
The subscript index $\alpha$(= $a, b, c$) stands for the orientation of an external magnetic field $B$ along the $\alpha$ axis. 
(1/$T_\mathrm{2L}$)$_{ab}$ is the relaxation rate for $B$$\perp$$c$ axis.  
Both (1/$T_\mathrm{2L}$)$_{cc}$ and (1/$T_\mathrm{2L}$)$_{ab}$ show an enhancement at 220$-$240 K,
which we call $T_2$ anomaly.  
To our knowledge, there has been no report on this $T_2$ anomaly in the pseudogap state.
 
For the central transition line ($I_z$ = 1/2 $\leftrightarrow$ -1/2) of a nuclear spin $I$ = 3/2, the transverse relaxation rate (1/$T_\mathrm{2L}$)$_{\gamma\gamma}$ is expressed by 
\begin{equation}
{\Bigl(\frac{1}{T_\mathrm{2L}}\Bigr)_{\gamma\gamma}} = {\frac{7}{2}}[J_{\alpha\alpha}(\nu_n) + J_{\beta\beta}(\nu_n)] + J_{\gamma\gamma}(0),
\label{eq2}
\end{equation}  
where $J_{\alpha\alpha}(\nu_n)$ ($\alpha, \beta, \gamma$ is the cyclic permutation of $a, b, c$) is the $\alpha$ component of the local filed fluctuations at the NMR frequency $\nu_n$~\cite{Moriya,Jac,Slichter,PHP,T2}.  
$J_{\alpha\alpha}(\nu_n)$ is expressed by the hyperfine coupling constants $A_{\alpha\alpha}$ and the electron spin-spin correlation function $S_{\alpha\alpha}$({\mbox{\boldmath $q$}}, $\nu$) 
\begin{equation}
J_{\alpha\alpha}(\nu_n) = \sum_{\mbox{\boldmath $q$}} A_{\alpha\alpha}({\mbox{\boldmath $q$}})^2 S_{\alpha\alpha}({\mbox{\boldmath $q$}}, \nu_n), 
\label{eq3}
\end{equation}   
where $S_{\alpha\alpha}$({\mbox{\boldmath $q$}}, $\nu$) is related to the dynamical spin susceptibility $\chi^{\prime\prime}$({\mbox{\boldmath $q$}}, $\omega$) through the fluctuation-dissipation theorem~\cite{Moriya,Jac,Slichter}.  
The nuclear spin-lattice relaxation rate (1/$T_1$)$_{\gamma\gamma}$ is expressed by 
\begin{equation}
{\Bigl(\frac{1}{T_\mathrm{1}}\Bigr)_{\gamma\gamma}} = J_{\alpha\alpha}(\nu_n) + J_{\beta\beta}(\nu_n).
\label{eq4}
\end{equation}
A conventional fluctuation spectrum is independent of the frequency in the NMR frequency region and then
$J_{\gamma\gamma}(0)$ = $J_{\gamma\gamma}(\nu_n)$ should hold.
If $J_{\gamma\gamma}(0)$ = $J_{\gamma\gamma}(\nu_n)$,
1/$T_{2L}$ should agree with the Redfield relaxation rate 1/$T_{2R}$ in Refs.~\cite{P,PHP,T2R} from the anisotropic $T_1$'s.
We estimate 1/$T_{2R}$ from
\begin{equation}
{\Bigl(\frac{1}{T_\mathrm{2R}}\Bigr)_{cc}} = 3{\Bigl(\frac{1}{T_{1}}\Bigr)_{cc}} + {\Bigl(\frac{1}{T_{1}}\Bigr)_{ab}},
\label{eq5}
\end{equation}
\begin{equation}
{\Bigl(\frac{1}{T_\mathrm{2R}}\Bigr)_{ab}} = {\frac{7}{2}}{\Bigl(\frac{1}{T_{1}}\Bigr)_{ab}} + {\frac{1}{2}}{\Bigl(\frac{1}{T_{1}}\Bigr)_{cc}},
\label{eq6}
\end{equation}  
according to uniaxially symmetric fluctuations ($J_{aa}$ = $J_{bb}$ is then denoted by $J_{ab}$)~\cite{P,PHP,T2R,Walstedt0,Walstedt,Narath}. 
(1/$T_\mathrm{2R}$)$_{ab}$ is the relaxation rate for $B$$\perp$$c$ axis. 
 
Figure~\ref{f4} (a) shows $^{63}$Cu nuclear spin-lattice relaxation rates $(1/T_{1})_{cc, ab}$ for the optimally doped Hg1212.
No frequency dependence was observed at 85$-$115 MHz. 
Figure~\ref{f4} (b) shows $^{63}$Cu nuclear spin-echo transverse relaxation rates $(1/T_\mathrm{2L})_{cc, ab}$ and the Redfield relaxation rates $(1/T_\mathrm{2R})_{cc, ab}$ estimated from the anisotropic $T_1$ using eqs.~(\ref{eq5}) and~(\ref{eq6}).   
In Fig.~\ref{f4} (b), the enhancement in $(1/T_\mathrm{2L})_{\gamma\gamma}$ at 220$-$240 K is not reproduced from $(1/T_\mathrm{2R})_{\gamma\gamma}$.  

Let us define $\Delta J_{\gamma\gamma}$ by
\begin{equation}
\Delta J_{\gamma\gamma} = {\Bigl(\frac{1}{T_\mathrm{2L}}\Bigr)_{\gamma\gamma}} - {\Bigl(\frac{1}{T_\mathrm{2R}}\Bigr)_{\gamma\gamma}}, 
\label{eq7}
\end{equation}
which can be a criterion to test whether the characteristic fluctuation frequency is higher or lower than the NMR frequency. 
Figure~\ref{f5} shows the temperature dependences of $\Delta J_{cc}$ and $\Delta J_{ab}$ for the optimally doped Hg1212. 
For $B$$\parallel$$c$ axis, $\Delta J_{cc}$ is equal to $J_{cc}(0) - J_{cc}(\nu_n)$.
Figure~\ref{f5} indicates $\Delta J_{cc}>$ 0 and $J_{cc}(0)$ strongly enhanced at 220$-$240 K.  
The characteristic frequency should be lower than the NMR frequency. 
Novel ultraslow fluctuations develope in the pseudogap state.
 
Isotope measurements could be helpful to specify which type the relaxation mechanism is magnetic or electric, because the isotope $^{65}$Cu has smaller quadrupole moment than $^{63}$Cu while the nuclear gyromagnetic ratio is larger. 
However, since our $T_\mathrm{2L}$ values of $^{65}$Cu were not conclusive due to the poor signal intensity, 
the effects of the charge fluctuations are unclear. 
 
\begin{figure}[h]
 \begin{center}
 \includegraphics[width=0.80\linewidth]{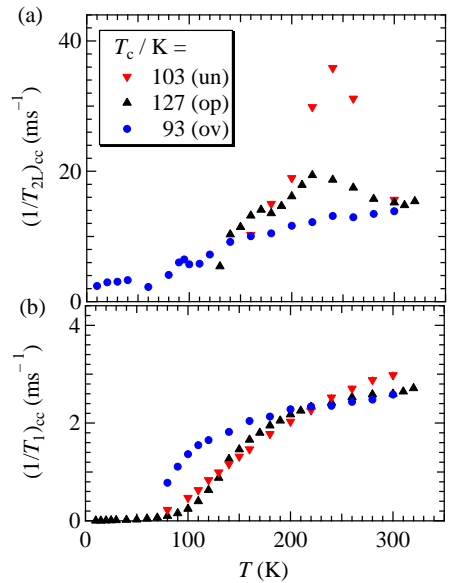}
 \end{center}
 \caption{\label{f6}(Color online)
(a) Doping dependence of (1/$T_\mathrm{2L}$)$_{cc}$ against temperature for Hg1212 from underdoped ($T_\mathrm{c}$ = 103 K) to overdoped ($T_\mathrm{c}$ = 93 K). 
(b) Doping dependence of (1/$T_\mathrm{1}$)$_{cc}$ against temperature for Hg1212 from underdoped ($T_\mathrm{c}$ = 103 K) to overdoped ($T_\mathrm{c}$ = 93 K).
}
 \end{figure} 
 
Figures~\ref{f6} (a) shows (1/$T_\mathrm{2L}$)$_{cc}$  against temperature for Hg1212 from underdoped ($T_\mathrm{c}$ = 103 K) to overdoped ($T_\mathrm{c}$ = 93 K). 
Figures~\ref{f6} (b) shows the doping dependence of (1/$T_\mathrm{1}$)$_{cc}$ against temperature.
The doping dependence of (1/$T_\mathrm{2L}$)$_{cc}$ is different from that of (1/$T_\mathrm{1}$)$_{cc}$. 
The peak temperature of (1/$T_\mathrm{2L}$)$_{cc}$ is nearly independent of the doping level,
but the enhancement in (1/$T_\mathrm{2L}$)$_{cc}$ is suppressed by overdoping. 
The peak temperature of (1/$T_\mathrm{2L}$)$_{cc}$ is similar to but deviates from the pseudo spin-gap temperature defined by the maximum of the $^{63}$Cu 1/$T_1T$~\cite{ItohHg1212}.

\section{Discussions}
As the carrier concentration increases, the pseudo spin-gap temperature of the $^{63}$Cu 1/$T_1T$ decreases and the peak value increases~\cite{ItohHg1212}, being in contrast to (1/$T_\mathrm{2L}$)$_{cc}$ in Fig.~\ref{f6} (a). 
Thus, the local field fluctuations causing $^{63}$Cu (1/$T_\mathrm{2L}$)$_{cc}$ at 220$-$240 K must be different from those causing $^{63}$Cu 1/$T_1T$.  

For Hg1212, we infer an ultraslow fluctuation spectrum of 
$\chi^{\prime\prime}_L(q, \omega)$/$\omega$ = $\chi_L(q\xi)\Gamma_L(q\xi)$/[$\omega^2$ + $\Gamma_L^2(q\xi)$] 
($\omega$ = 2$\pi\nu$) with a characteristic energy scale $\Gamma_L(q\xi)$
[ 50 kHz ${<}$ $\Gamma_L/2\pi$ ${<}$ $\nu_n$: an ultraslow condition ]
and a correlation length $\xi$,
in addition to the antiferromagnetic spin fluctuation model~\cite{BS,MTU,MMP}. 
Then, we obtain 1/$T_1$ ${\ll}$ $\chi_L/\Gamma_L$ $\sim$ 1/$T_\mathrm{2L}$, which shows a peak at 220$-$240 K for Hg1212.   

The emergence of the ultraslow fluctuations has been reported near the charge-spin stripe ordering temperature and the phase boundary in the La-based 214 systems (LSCO)~\cite{LSCO1,LSCO2,LSCO3,LSCO4}.
The wipeout effect on NMR/NQR signals is also characteristic of LSCO~\cite{LSCO1,LSCO2,LSCO3}. 
Hg1212, however, shows no wipeout effect on the Cu NMR spectrum nor any glassy behavior in the pseudogap state. 
The $T_2$ anomaly of Hg1212 is a homogeneous phenomena.  

In La$_{1.6-x}$Nd$_{0.40}$Sr$_{x}$CuO$_4$, 
the transverse relaxation rate 1/$^{139}T_2$ of the $^{139}$La nuclear spin shows a peak at about 20 K above $T_c$ and below the charge ordering temperature $T_{charge}$$\sim$70 K ~\cite{LSCO1}, and 1/$^{63}T_\mathrm{2L}$ of the $^{63}$Cu nuclear spin shows a peak at about 15 K or 50 K~\cite{LSCO2}.  
In La$_{1.68}$Eu$_{0.20}$Sr$_{0.12}$CuO$_4$, the $^{139}$La nuclear spin-lattice relaxation rate 1/$^{139}T_1$ also shows a peak above $T_c$ and below $T_{charge}$, while the $^{63}$Cu 1/$^{63}T_1$ shows a weak upturn on cooling below $T_{charge}$~\cite{LSCO2,LSCO3}.
The $^{139}$La nuclear spin play a role in detecting lower frequency fluctuations than the $^{63}$Cu nuclear spin. 
The difference in the relaxation rates between $^{139}$La and $^{63}$Cu for (Nd, Eu)-doped LSCO is parallel to that between the $^{63}$Cu 1/$^{63}T_\mathrm{2L}$ and 1/$^{63}T_\mathrm{1}$ for Hg1212 in Fig.~\ref{f6}. 
For the time being, it is unlikely that Hg1212 is in a charge ordering state at 220$-$240 K, because the high electrical conductivity shows a metallic behavior~\cite{R1,R2}. 
The ultraslow fluctuation spectrum $\chi^{\prime\prime}_L(q, \omega)$/$\omega$ of Hg1212 may be a part of the charge-spin stripe fluctuation spectrum but should describe a purely fluctuating stripe. 

Recent experimental efforts have revealed several hidden orders in the pseudogap states:
quadrupolar fluctuations below 200 K in YBa$_2$Cu$_4$O$_8$~\cite{CDWY1248}, 
the onset $T_\mathrm{CDW}$ of a short-ranged charge density wave~\cite{CDW,CDM}, the development of an intra-unit-cell local magnetic order~\cite{Sidis}, and the onset of finite Kerr rotations~\cite{Kerr}.   
The $T_2$ anomaly of Hg1212 may correspond to one of these short-ranged orders, because (1/$T_\mathrm{2L}$)$_{cc}$ shows a peak but not divergence. 
These findings suggest that the pseudogap state is not a simple spin-singlet state.

\section{Conclusions}
In conclusion, we found the enhanced zero frequency local field fluctuations at 220$-$240 K for Hg1212 in the pseudogap states through the $^{63}$Cu nuclear spin-echo transverse relaxation, which suggests the emergence of ultraslow fluctuations. 
Novel ultraslow fluctuations without any glassy behavior are hidden in the pseudogap states. 




\begin{thebibliography}{50}
\bibitem{T2}To be more precise, a local field fluctuation in 1/$T_\mathrm{2R}$ of the central transition line is also a function of the Larmor frequencies of $I_z$ = $\pm$3/2 $\leftrightarrow$ $\pm$1/2.
\end{thebibliography}
\end{document}